\documentclass[12pt,a4paper,dvips]{article}
\usepackage{psfig}
\usepackage{epsfig}
\renewcommand\thesection{\arabic{section}.}
\renewcommand\thesubsection{\thesection\arabic{subsection}.}
\renewcommand\thesubsubsection{\thesubsection\arabic{subsubsection}.}
\renewcommand\section[1]{\vspace{\topsep}\vspace{\partopsep}
\refstepcounter{section}
{\par  \noindent\normalsize\bfseries \thesection
\hspace{1em}#1\vspace{\topsep}\par\noindent}}

\newenvironment{refs}
{\vspace{\topsep}\vspace{\partopsep}
{\par \noindent\normalsize\bfseries  References
\vspace{-\topsep}\par\noindent}
\setlength{\parindent}{-5mm}
\begin{list}{}{\topsep 0pt \partopsep 0pt \itemsep 0pt \leftmargin 5mm
\parsep 0pt \itemindent -5mm}}
{\end{list}}

\renewcommand\subsection[1]{
\refstepcounter{subsection}
{\par \protect\vspace{\topsep}\vspace{\partopsep}
 \noindent\normalsize\bfseries \slshape \thesubsection
\hspace{1em}#1\par \noindent}}

\renewcommand\subsubsection[1]{
\refstepcounter{subsubsection}
{\par \protect \vspace{\topsep}\vspace{\partopsep}
\noindent\normalsize \slshape \thesubsubsection
\hspace{1em}#1\par \noindent}}

\newfont{\sansb}{cmssbx10}
\newfont{\sans}{cmss10}
\begin{document}
\begin{center}
{\large \bf Spectrum and Variability of Mrk501 as observed by the CAT
Imaging Telescope \vspace{18pt}\\}
{A. Djannati-Ata\"\i$^1$  for the CAT Collaboration \vspace{12pt}\\}
{\sl 
$^1$LPNHE Paris University 6\&7, 4 Place Jussieu - T33 RdC, 75252
Paris Cedex 05, France\\
}
\end{center}
%\section*{\center \vspace{-24pt}Abstract}
\begin{abstract}

The C{\small AT} Imaging Telescope has observed the BL Lac object Markarian 501
between March and 
August 1997. We report here on the variability over this time
including several large flares. We present also preliminary spectra
for all these data, for the low emission
state, and for the largest flare. 
\end{abstract}

\setlength{\parindent}{1cm}
\section{Introduction and Data Sample}
Several groups have reported strong emission from Mrk501 at this
workshop.
The C{\small AT} Imaging Telescope, operating since October 1996 and (described
in Rivoal et al. 1997), is located 
in the French Pyrenees (42,5$^\circ$ N, 2$^\circ$ E) at an altitude of
1650~m. It has a mirror area of 17~m$^2$ with a fine grained imaging
camera (546 pixels, each 0.12$^\circ$ for these data) and a threshold
energy of 220 GeV.
Between March and August 1997 a total of 80 hours of data were taken
on Mrk501 with 25 hours of control region (OFF) data.
As the performance of the telescope at large zenith angles is still
under investigation, the data used for the variability study were
limited to those at less than 25$^\circ$ from Zenith, and
for the spectral analysis at less than 10$^\circ$ (16.58 hours).

\section{Analysis method}
A standard moment-based (``SuperCuts'') analysis has been used for the
study of Mrk501's light curve (section 3).
The cuts used are lower than for other detectors, taking advantage of
the C{\small AT} telescope's very high resolution imaging camera, as follows:
\vspace{-3mm}
\begin{center}
\begin{tabular}{cccccl}
 0.7 & $<$ &  Width &  $<$ & 1.5 & (mrad) \\

 2.0 & $<$ & Length &  $<$ & 5.0  & (mrad) \\

 2.0 &  $<$ & Distance &   &   & (mrad)\\

 30  &  $<$ &  Size    &   &   &  (p.e.)\\

   &  &   $\alpha$  &  $<$ & 9  &   (deg.)\\
\end{tabular}
\end{center}

\vspace{-2mm}
A new method has been developed for the C{\small AT}
telescope (Lebohec 1996 \& Lebohec et al. 1995) which provides directly the energy
of the incident $\gamma$-ray; this has been used in the spectral analysis.
The method uses a
maximum-likelihood fit to the images of an analytical model for 
$\gamma$-shower images. The image is given by the model as a function
of primary energy, impact parameter (R), angular source position in
the field of view and azimuthal position of the image about the source.
Minimization of the $\chi^2$ fit provides an estimate of these
parameters. In addition, the minimized $\chi^2$ value gives a
criterion for selection of good $\gamma$-ray candidates.
The orientation angle $\alpha$ in this method is defined (between 0
and 180$^\circ$) as the angle at the image centre between the expected
source position and reconstructed source position. 
Figure 1 shows the clear signal in the alpha plot of the data from Mrk501 at 
less than 25$^\circ$ from Zenith. 

\begin{figure}[htb]
\centerline{\epsfig{file=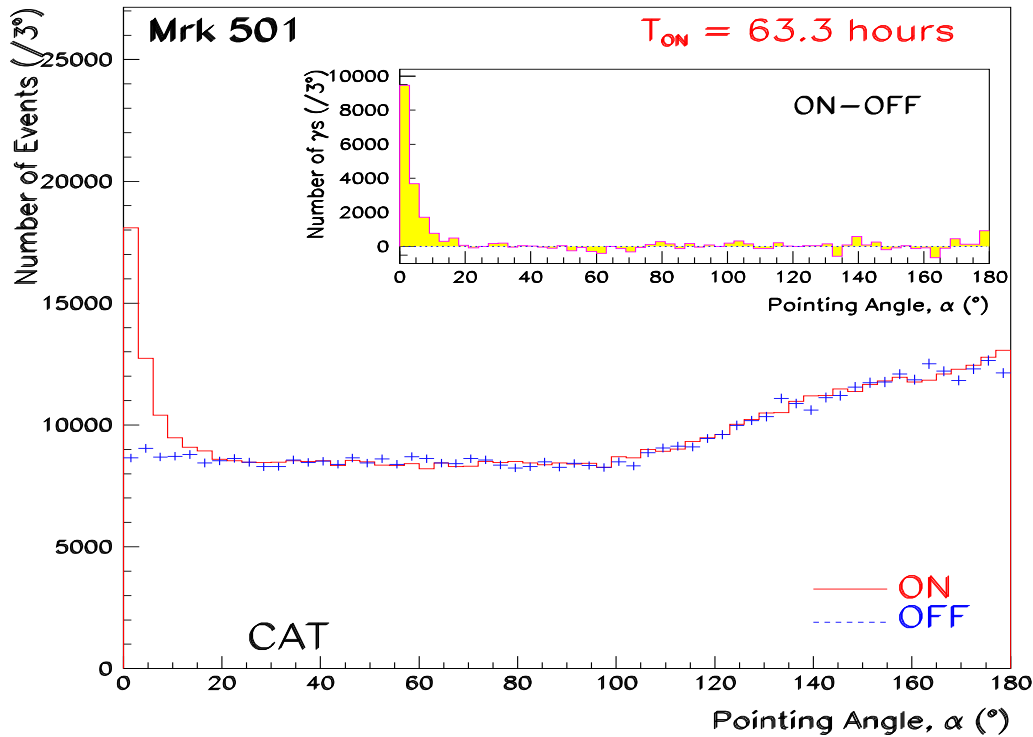,width=14cm,height=11cm}}
\caption{ On-source and Off-source $\alpha$-distributions 
for March to August observations of Mrk501.  }
\end{figure}

The origin of each gamma shower on the celestial sphere can be reconstructed 
event by event to 0.16$^\circ$. 
The energy resolution obtained with this method is 20\%, independent
of energy from 200~GeV to 10~TeV.

Three cuts are
used for the spectral analysis (derived from simulations to optimize
the number of $\gamma$-ray candidates):
\vspace{-3mm}
\begin{center}
\begin{tabular}{cccccl}

 0.2  &$<$ &    $P_{\chi^2}$ & &  &\\

 40.0 &$<$ &    R   &         &  &  (metres)	\\

 & &   $\alpha$ &   $<$ &  9     & (deg.)\\

\end{tabular}
\end{center}

\vspace{0.5cm}
\section{Source Variability and Light Curve}
The nightly flux levels (expressed in gamma-rays per minute
uncorrected for cut efficiencies) for Mrk501
between March 7 and August 8 1997 are shown in figure 2. 
During this period the average flux was higher than that of the 
Crab Nebula, with the highest flare over 6 times greater.
Variability on a day time-scale is evident in this figure;
shorter time variations are still under study.

\begin{figure}[htb]
\centerline{\epsfig{file=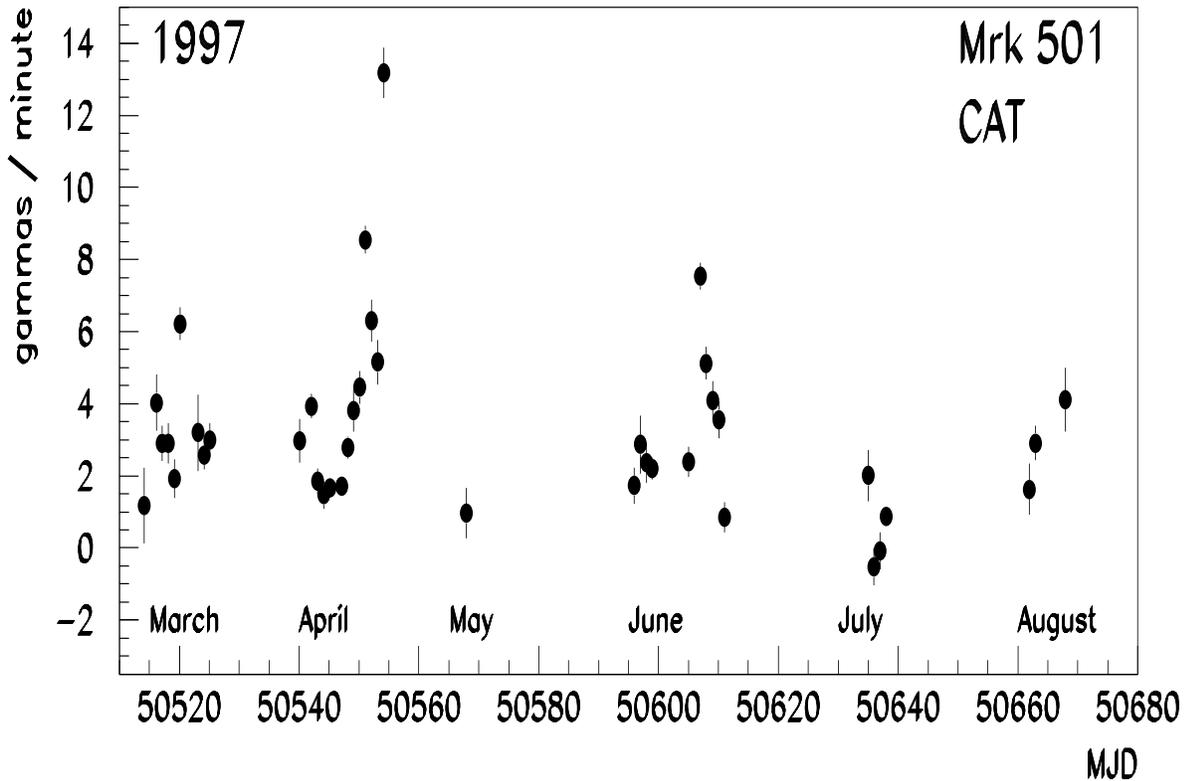,width=16cm,height=11cm}}
\caption{Nightly gamma-ray rate for Mrk501 from March to August, 1997.
Coverage of
the source was reduced in May due to technical 
difficulties, and in August due to weather conditions.}
\end{figure}

Figure 3 shows the remarkable agreement between the data from three
observatories (C{\small AT}, Whipple, and H{\small EGRA}) for the April period. 
The data
from each experiment have been rescaled as they have different
thresholds and observatory altitudes.  This demonstrates also the
utility/necessity of having several observatories to follow
continuously such variable sources. 

\begin{figure}[htb]
\centerline{\epsfig{file=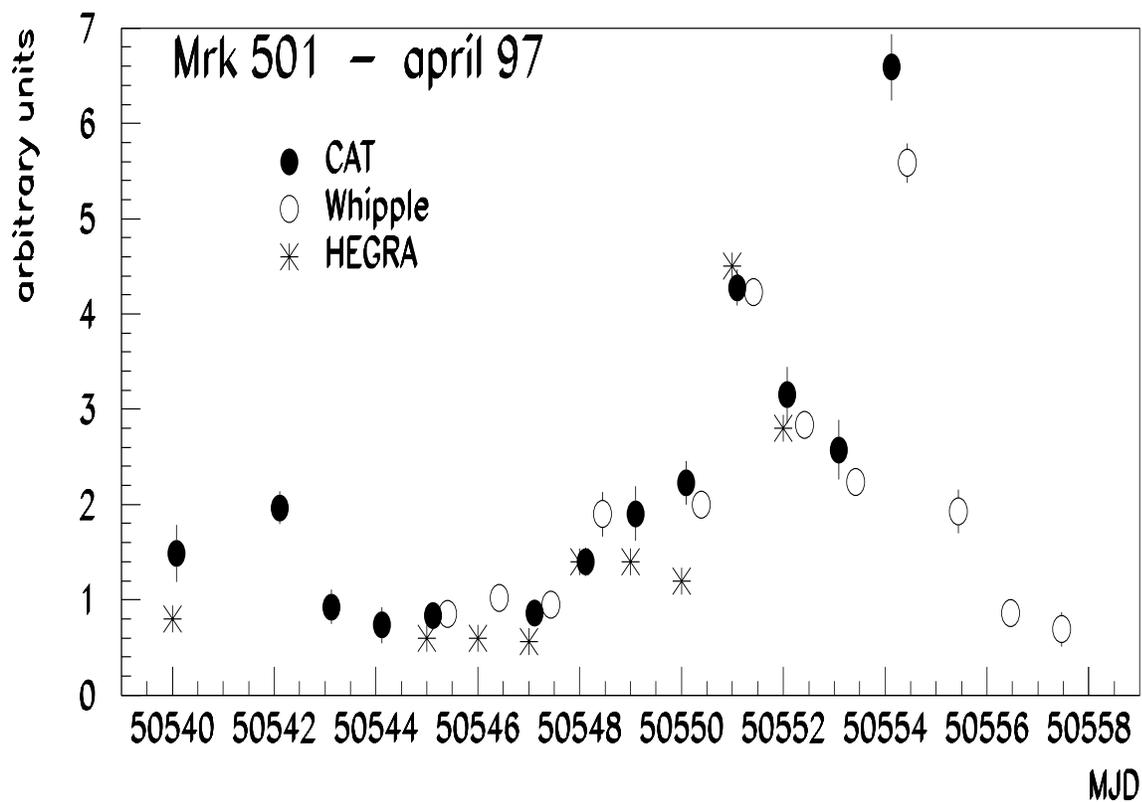,width=16cm,height=11cm}}
\caption{April 1997 nightly gamma-ray rate for Mrk501 as seen by three
groups : C{\small AT}, Whipple (Fegan, 1997), and H{\small EGRA} system
(Lorenz, 1997) . }
\end{figure}

\section{Differential Energy Spectra}
The acceptance area and the cut efficiency were derived from detailed
Monte Carlo simulations as a function of energy and zenith angle. 
These, combined with the energies from the maximum-likelihood fit, allow
differential energy spectra to be obtained. Spectra from Mrk501 are
shown in figure 4 (for data less than 10$^\circ$ from Zenith): for all
data, for the data for which the source was in a low state, and for
the largest flare on the night of April 15-16th. A maximum-likelihood
fit of a power law to the spectrum of these data gives for all data:
\begin{equation}
\frac{dN}{dE} = 5.89 \pm 0.22 \times 10^{-7} E_{TeV}^{-2.40 \pm 0.04}
m^{-2}s^{-1}TeV^{-1} 
\end{equation}
The corresponding flux and index for the low state data are
$2.50 \pm 0.18$ and $-2.59 \pm 0.09$, and for the largest flare
$18.94 \pm 0.64$ and $-2.33 \pm 0.05$.

The errors in these formula are statistical only as the systematics
studies are underway. Within statistical errors the spectrum index seems
compatible with a constant hardness of the source.  
A cross-check of these values by the standard SuperCuts 
analysis yields closely compatible results both for flux values and spectral
indices.   The spectra appear to be harder than that of the Crab Nebula
(given in Goret et al., 1997).  In this dataset alone, there is a 4.7$\sigma$
signal at energies above 7~TeV; if the cut on  $P_{\chi^2}$ is relaxed the 
signal increases to 6.3$\sigma$ above 7~TeV, with 4.8$\sigma$ above 10~TeV.
Further studies using the data taken at high 
Zenith angle should extend the spectral coverage to higher energies.

\begin{figure}[htb]
\centerline{\epsfig{file=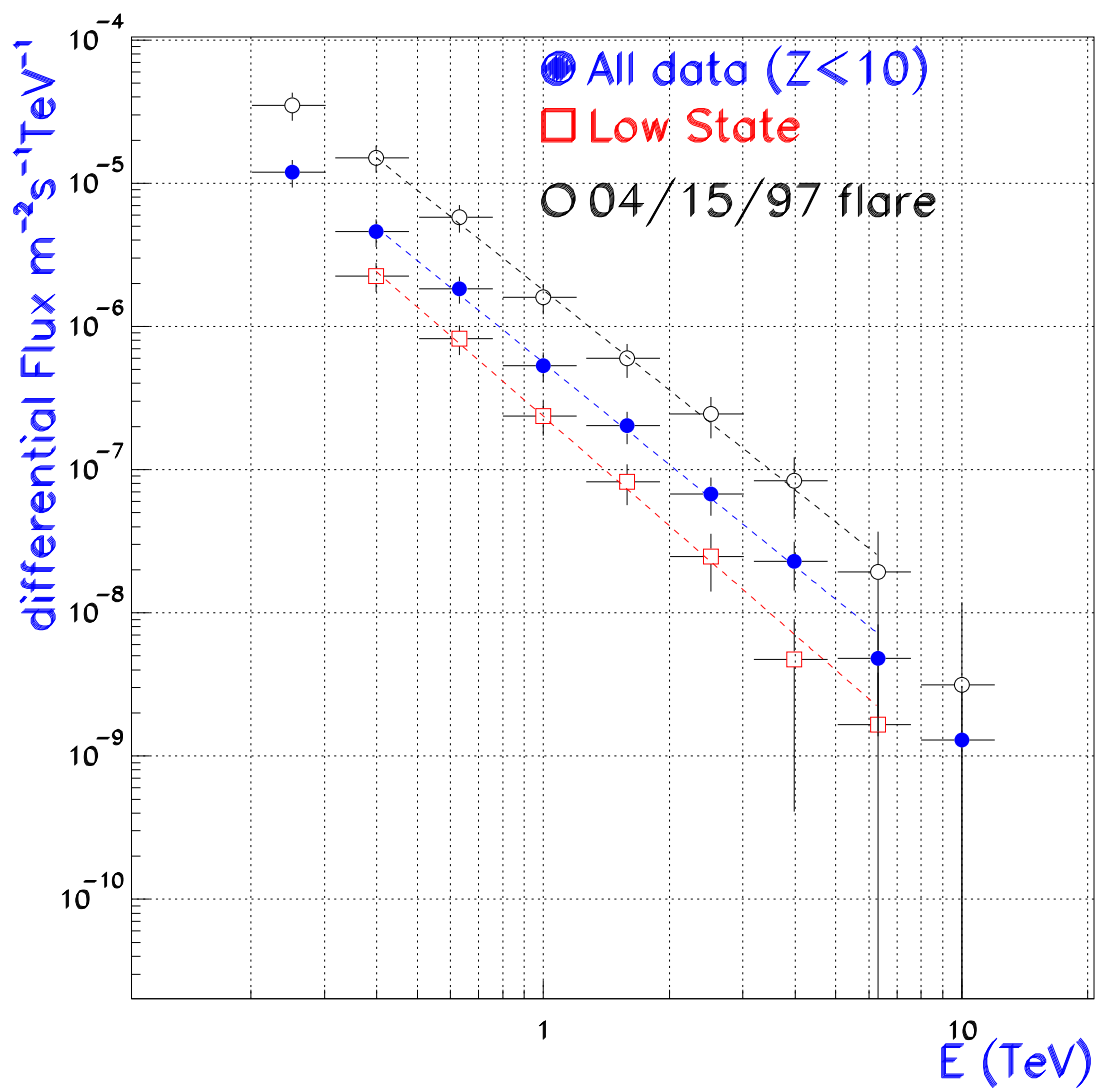,width=16cm,height=16cm}}
\caption{ Differential spectra for Mrk501 for all data, low-state emission,
and the highest flare.}
\end{figure} 

\section{Conclusions}
The source Markarian 501 entered a very active phase in 1997, with several
large flares during which it became the brightest object in the VHE 
$\gamma$-ray sky.  This unusual behaviour has been detected by the C{\small AT}
imaging telescope and several other groups.  The intensity of the
emission provides data which are exteremely
rich in $\gamma$-rays, useful both for testing of the detector capabilities and
investigation of the highest-energy part of the spectrum.
At the time of writing the source has been seen to be still active in
data taken at high Zenith angle.

%\section{Acknowledgements}
%This section is optional.

\begin{refs}
\item Rivoal R.M., et al., 1997, Proc. 25th ICRC, 5, 89
\item Lebohec, S., 1996, Doctoral Thesis
\item Lebohec, S., et al., 1995, Proc. ``Towards a Major Atmospheric \v Cerenkov
Detector IV'', ed. M. Cresti , 378
\item Fegan, D.J., 1997, private communication
\item Lorenz, E., 1997, private communication
\item Goret, P., 1997, this workshop

\end{refs}

%\begin{table}[hbt]
%\center
%This sample manuscript in LaTeX is available at:
%http://fskbhe1.puk.ac.za/knpc/style
%\end{table}
\end{document}